\documentclass[12pt]{iopart}
\usepackage[dvips]{graphicx}
\usepackage{iopams}
\usepackage{cite}
\begin{document}
\title[Stationary two-atom entanglement]{Stationary two-atom entanglement 
induced by nonclassical two-photon correlations}
\author{R Tana\'s\dag\ and Z Ficek\ddag\ } 
\address{\ddag\ Nonlinear Optics Division, Institute of Physics, 
Adam Mickiewicz University, Pozna\'n, Poland\\
\ddag Department of Physics, School of Physical Sciences, The
University of Queensland, Brisbane, QLD 4072, Australia}

\ead{tanas@kielich.amu.edu.pl}

\begin{abstract}
A system of two two-level atoms interacting with a squeezed vacuum 
field can exhibit stationary entanglement associated with nonclassical 
two-photon correlations characteristic of the squeezed vacuum field.
The amount of entanglement present in the system is quantified by
the well known measure of entanglement called concurrence.
We find analytical formulas describing the concurrence for two
identical and nonidentical atoms and show that it is possible to
obtain a large degree of steady-state entanglement in the
system. Necessary conditions for the entanglement are nonclassical 
two-photon correlations and nonzero collective decay.
It is shown that nonidentical atoms are a better source of stationary 
entanglement than identical atoms. We discuss the optimal physical
conditions for creating entanglement in the system, in particular, it
is shown that there is an optimal and rather small value of the mean
photon number required for creating entanglement.
\end{abstract}

\submitto{\JOB}
\maketitle

\section{Introduction}
Entanglement between separate quantum systems is one of the key 
problems in quantum mechanics. A number of interesting concepts and 
methods for creating entanglement have been proposed involving trapped and 
cooled ions or neutral atoms~\cite{bbtk,koz,hal,flei,sm01,b2,nat1,nat3}.
Of particular interest is generation of
entangled states in two-atom systems, since they can represent two
qubits, the building blocks of the quantum gates that are
essential to implement quantum protocols in quantum information processing.
It has been shown that entangled states in a two-atom system can be
created by a continuous driving of the atoms with a coherent
or chaotic thermal field~\cite{sm01,afs,klak,zsl,mfw}, or by a pulse
excitation followed by a continuous observation of radiative
decay~\cite{phbk,b1,cabr}. Moreover, the effect of spontaneous emission
on initially prepared entangled state has also been
discussed~\cite{gy1,gy2,bash,jak}. These studies,
however, have been limited to the small sample (Dicke) model~\cite{dic}
or the situation involving noninteracting atoms strongly coupled to
a cavity mode. The difficulty of the Dicke model is that it does not
include the dipole-dipole interaction among the atoms and does not
correspond to realistic experimental situations of atoms located
(trapped) at different positions. In fact, the model corresponds to a
very specific geometrical configuration of the atoms confined to a
volume much smaller compared with the atomic resonant wavelength (the
small-sample model). The present atom trapping and cooling
techniques can trap two atoms at distances of order of a resonant
wavelength~\cite{eich,deb,tos}, which makes questionable the
applicability of the Dicke model to physical systems.

Recently, we have shown~\cite{ft03} that spontaneous
emission from two spatially separated atoms can lead to a transient
entanglement of initially unentangled atoms. This result contrasts
with the Dicke model where spontaneous emission cannot produce entanglement
from initially unentangled atoms~\cite{klak,bash}.
We have also found~\cite{tf03} analytical results for two measures of
entanglement and the relation between them for the two-atom system
radiating by spontaneous emission for quite broad range of initial
conditions. 

In this paper we study the creation of a stationary entanglement in 
a system of two identical as well as nonidentical two-level atoms 
separated by an arbitrary distance $r_{12}$ and interacting with a 
squeezed vacuum. The squeezed vacuum appears here as a source of 
nonclassical two-photon coherences, essential for the creation of the 
stationary entanglement.
We use the master equation to describe the evolution of the system and
find the steady-state solutions for the atomic variables.
We present analytical results for {\em concurrence} which is well known
and calculable measure of entanglement. We find a surprising result 
that non-identical atoms with significantly different transition 
frequencies can exhibit a larger entanglement than identical atoms. 
Under some conditions, the nonidentical atoms can be maximally 
entangled with the value of the concurrence equal to unity.

\section{Master equation}
We consider a system of two two-level atoms at fixed positions 
$\mathbf{r}_{1}$ and $\mathbf{r}_{2}$ and coupled to the radiation
field, whose the modes are in a squeezed vacuum state. Each atom has 
energy levels $\left|g_{i}\right\rangle$ and 
$\left|e_{i}\right\rangle \ (i=1,2)$ such that $E_{e_{i}}-E_{g_{i}} 
=\hbar \omega_{i}$, transition dipole moment $\boldsymbol{\mu}_{i}$, 
which we assume equal for both atoms.

We analyse separately the dynamics of identical and non-identical 
atoms. In the case of nonidentical atoms, we assume different 
transition frequencies $\omega_{1}$ and
$\omega_{2}$ such that $\Delta =(\omega_{2}-\omega_{1})/2 \ll \omega_{0}=
(\omega_{1}+\omega_{2})/2$, so that the rotating-wave approximation
can be applied to calculate the dynamics of the system.

The system can be described by the reduced density operator $\rho$ 
which, in a Schr\"odinger picture, satisfies the master 
equation~\cite{ft02} 
\begin{eqnarray}
\frac{\partial {\rho} }{\partial t} &=&
-\frac{1}{2}\sum_{i,j=1}^{2}\Gamma _{ij}
\left( 1+N \right) \left( {\rho}
S_{i}^{+}S_{j}^{-}+S_{i}^{+}S_{j}^{-}{\rho}
-2S_{j}^{-}{\rho} S_{i}^{+}\right) \nonumber \\
&&-\frac{1}{2}\sum_{i,j=1}^{2}\Gamma _{ij}N
\left( {\rho} S_{i}^{-}S_{j}^{+}+S_{i}^{-}S_{j}^{+}{\rho}
-2S_{j}^{+}{\rho} S_{i}^{-}\right) \nonumber \\
&&+\frac{1}{2}\sum_{i,j=1}^{2}\Gamma _{ij}M
\left( {\rho} S_{i}^{+}S_{j}^{+}+S_{i}^{+}S_{j}^{+}{\rho}
-2S_{j}^{+}{\rho} S_{i}^{+}\right)e^{-2i\omega_{s}t} \nonumber \\
&&+\frac{1}{2}\sum_{i,j=1}^{2}\Gamma _{ij}
M^{\ast }\left( {\rho} S_{i}^{-}S_{j}^{-}+S_{i}^{-}S_{j}^{-}{\rho}
-2S_{j}^{-}{\rho} S_{i}^{-}\right)e^{2i\omega_{s}t} \nonumber \\
&&-i\sum_{i=1}^{2}\omega_{i}\left[S^{z}_{i},{\rho} \right]
-i\sum_{i\neq j}^{2}\Omega_{ij}\left[ S^{+}_{i}S^{-}_{j},
{\rho} \right] .  \label{eq1}
\end{eqnarray}
Here, $S_{i}^{+}$ and $S_{i}^{-}$ are the raising and lowering 
operators, respectively, of the $i$th atom, 
$N$ and $M=|M|\exp\left(i\phi_{s}\right)$ characterise squeezing 
such that $|M|^{2}\leq N(N+1)$, where the equality holds for a 
minimum-uncertainty squeezed state, $\phi_{s}$ is the squeezing phase and 
$\omega_{s}$ is the carrier frequency of the squeezed vacuum.

The parameters $\Gamma_{ij}$, which appear in equation~(\ref{eq1}),
are spontaneous emission rates, such that
\begin{eqnarray}
      \Gamma_{ii}\equiv \Gamma =
      \frac{\omega_{0}^{3}|\boldsymbol{\mu}|^{2}}{3\pi \varepsilon_{o}\hbar
      c^{3}},\quad (i=1,2) \label{eq2}
\end{eqnarray}
is the spontaneous emission rate of the $i$th atom, assumed to be equal 
for both atoms, and
\begin{eqnarray}
\Gamma_{12}=\Gamma_{21}&=&
\frac{3}{2}\sqrt{\Gamma}
\left\{ \left[1 -\left( \hat{\boldsymbol{\mu}}\cdot
      \hat{\mathbf{r}} 
_{12}\right)^{2} \right] \frac{\sin \left( k_{0}r_{12}\right)
}{k_{0}r_{12}}\right.\nonumber\\
&&\left. +\left[ 1 -3\left( \hat{\boldsymbol{\mu}}\cdot
\hat{\mathbf{r}}_{12}\right)^{2} \right] \left[ \frac{\cos \left(
k_{0}r_{12}\right) }{\left( k_{0}r_{12}\right) ^{2}}-\frac{\sin \left(
k_{0}r_{12}\right) }{\left( k_{0}r_{12}\right) ^{3}}\right] \right\} \ ,
\label{t33}
\end{eqnarray}
are collective spontaneous emission rates arising from the coupling
between the atoms through the vacuum
field~\cite{ftk87,leh,ag74}, and
\begin{eqnarray}
\Omega_{12} &=&\frac{3}{4}\,\Gamma\left\{ -\left[
1-\left( \hat{\boldsymbol{\mu}}\cdot \hat{\mathbf{r}}%
_{12}\right)^{2} \right] \frac{\cos \left( k_{0}r_{12}\right)
}{k_{0}r_{12}}\right.  \nonumber \\
&&\left. +\left[ 1 -3\left( \hat{\boldsymbol{\mu}}\cdot
\hat{{\mathbf{r}}}_{ij}\right)^{2} \right] \left[ \frac{\sin \left(
k_{0}r_{12}\right) }{\left( k_{0}r_{12}\right) ^{2}}+\frac{\cos \left(
k_{0}r_{12}\right) }{\left( k_{0}r_{12}\right) ^{3}}\right] \right\}\, .
\label{t39}
\end{eqnarray}
represents the vacuum induced coherent (dipole-dipole) interaction
between the atoms. 

In the expressions~(\ref{t33}) and~(\ref{t39}),
$\hat{\boldsymbol{\mu}}$ and $\hat{\mathbf{r}}_{12}$ are unit vectors 
along the atomic transition dipole moments and the vector
$\mathbf{r}_{12}=\mathbf{r}_{2}-\mathbf{r}_{1}$, 
respectively, and $k_{0}=\omega_{0}/c$. Later on, we will assume that
the atomic dipole moments ${\boldsymbol{\mu}}$ are perpendicular
to the vector $\mathbf{r}_{12}$ joining the two atoms.

The collective parameters $\Gamma_{12}$ and
$\Omega_{12}$, which arise from the mutual interaction between the atoms,
significantly modify the master equation of a two-atom system. The parameter
$\Gamma_{12}$ introduces a coupling between the atoms through the vacuum
field that the spontaneous emission from one of the atoms influences the
spontaneous emission from the other. The dipole-dipole interaction
term $\Omega_{12}$ introduces a coherent coupling between the atoms. Owing
to the dipole-dipole interaction, the population is coherently
transferred back and forth from one atom to the other.

The two-atom system can be described in the basis of product states of
the individual atoms 
\begin{eqnarray}
  \label{eq:basis}
|1\rangle=|g_{1}\rangle\otimes|g_{2}\rangle=|g\rangle\, ,\nonumber\\
|2\rangle=|e_{1}\rangle\otimes|e_{2}\rangle=|e\rangle\, ,\nonumber\\
|3\rangle=|g_{1}\rangle\otimes|e_{2}\rangle\, ,\nonumber\\
|4\rangle=|e_{1}\rangle\otimes|g_{2}\rangle\, ,
\end{eqnarray}
where $|g_{i}\rangle$ and $|e_{i}\rangle$ (for $i=1,2$) are the ground
and excited states of the individual atoms. In this basis, the 
two-atom system behaves as a single four-level system whose the density
matrix can be written as a 4$\times$4 matrix. Due to the presence of
the dipole-dipole interaction $\Omega_{12}$ it is often convenient to
introduce collective atomic states that are eigenstates of the system
of two identical atoms including the dipole-dipole 
interaction~\cite{dic}. They
are symmetric and antisymmetric superpositions of the product atomic
states $|3\rangle$ and $|4\rangle$, given by
\begin{eqnarray}
  \label{eq:collectivest}
|s\rangle &=&\phantom{-}\frac{1}{\sqrt{2}}\left(|3\rangle  
+|4\rangle\right)\, ,\nonumber\\
|a\rangle &=&-\frac{1}{\sqrt{2}}\left(|3\rangle -|4\rangle\right)\, ,
\end{eqnarray}
and the states $|g\rangle$ and $|e\rangle$ remain unchanged.

On introducing the collective states~(\ref{eq:collectivest}) and using
the master equation~(\ref{eq1}) we are able to write down equations of
motion for the matrix elements of the density matrix for two atoms in
a squeezed vacuum. We treat separately two cases of identical and
nonidentical atoms.

\subsection{Identical atoms}
For identical atoms separated by an arbitrary distance $r_{12}$ and
interacting with a squeezed vacuum field of the carrier frequency 
$\omega_{s}=\omega_{0}$, we obtain the following set of coupled equations
of motion~\cite{ft02}
\begin{eqnarray}
\dot{\rho}_{ee} &=&
-2\Gamma\left({N}+1\right)\rho_{ee}
+{N}\left[\left(\Gamma
+\Gamma_{12}\right)\rho_{ss} +\left(\Gamma
-\Gamma_{12}\right)\rho_{aa}\right]
+\Gamma_{12}|{M}|\rho_{u}\, ,\nonumber \\
\dot{\rho}_{ss} &=&
\left(\Gamma +\Gamma_{12}\right)\left\{{N}
-\left(3{N}+1\right)\rho_{ss} -{N}\rho_{aa}
+\rho_{ee} -|{M}|\rho_{u}\right\}\, ,\nonumber \\
\dot{\rho}_{aa} &=&
\left(\Gamma -\Gamma_{12}\right)\left\{{N}
-\left(3{N}+1\right)\rho_{aa} -{N}\rho_{ss}
+\rho_{ee}
+|{M}|\rho_{u}\right\}\, ,\nonumber \\
\dot{\rho}_{u} &=& 2\Gamma_{12}|{M}|
-\left(2{N}+1\right)\Gamma \rho_{u}
 -2|{M}|\left[\left(\Gamma
+2\Gamma_{12}\right)\rho_{ss} -\left(\Gamma
-2\Gamma_{12}\right)\rho_{aa}\right] \, ,\label{eq7}
\end{eqnarray}
where 
$\rho_{u}=\rho_{eg}\exp (-i\phi_{s}) +\rho_{ge}\exp (i\phi_{s})$. 
It is seen from equation~(\ref{eq7}) that the evolution of the populations
depends on the two-photon coherences $\rho_{eg}$ and $\rho_{ge}$,
which can modify population distribution between the collective states. 
The coherences can also create superposition (entangled) states 
involving only the ground $|g\rangle$ and the upper $|e\rangle$ states.
The evolution of the populations depends
on $\Gamma_{12}$, but is completely independent of the dipole-dipole
interaction $\Omega_{12}$.

The steady state solutions of equations (\ref{eq7}) depend on whether 
$\Gamma_{12}=\Gamma$ or $\Gamma_{12}\neq \Gamma$. 
For two atoms separated by an arbitrary distance $r_{12}$,
$\Gamma_{12}\neq \Gamma$, and then the steady-state
solutions of equations~(\ref{eq7}) are
\begin{eqnarray}
\rho_{ee}
&=&\frac{{N}^{2}\left[(2N+1)^{2}-4|M|^2\right]+|M|^{2}\gamma_{12}^{2}} 
{(2N+1)^{4}-4|M|^2\left[(2N+1)^2-\gamma_{12}^{2}\right]}\, ,\nonumber \\
\rho_{ss}&=&\frac{N(N+1)\left[(2N+1)^{2}-4|M|^2\right]
+|M|^{2}\gamma_{12}(\gamma_{12}-2)} 
{(2N+1)^{4}-4|M|^2\left[(2N+1)^2-\gamma_{12}^{2}\right]}\, ,  \nonumber \\
\rho_{aa} &=&\frac{N(N+1)\left[(2N+1)^{2}-4|M|^2\right]
+|M|^{2}\gamma_{12}(\gamma_{12}+2)}
{(2N+1)^{4}-4|M|^2\left[(2N+1)^2-\gamma_{12}^{2}\right]}\, ,  \nonumber \\
\rho_{u} &=&\frac{2\left(2{N}+1\right)|M|\gamma_{12}}{(2N+1)^{4}-4|M|^2
\left[(2N+1)^2-\gamma_{12}^{2}\right]}\, ,\label{t220}
\end{eqnarray}
where $\gamma_{12}=\Gamma_{12}/\Gamma$ is the dimensionless 
collective damping parameter.
This result shows that all the collective states are populated in the
steady-state even for small interatomic separations
$(\gamma_{12}\approx 1)$. For 
large interatomic separations $\gamma_{12}\approx 0$, and 
then the symmetric
and antisymmetric states are equally populated. When the interatomic
separation decreases, the population of the state $|a\rangle$ increases,
whereas the population of the state $|s\rangle$ decreases and
$\rho_{ss}=0$ for very small interatomic separations. This effect 
results from the enhanced $(\Gamma +\Gamma_{12})$ damping rate of the 
symmetric state, as it is seen from equation~(\ref{eq7}).
The two-photon coherences, represented by $\rho_{u}$, affect the population 
distribution only when both $|M|$
and $\gamma_{12}$ are nonzero. The coherences are crucial for getting
entanglement in the system. Of particular interest are population 
distributions for maximally squeezed fields with
$|M|=\sqrt{N(N+1)}$. In this case, the factor $(2N+1)^{2}-4|M|^{2}=1$, 
and then the solutions~(\ref{t220}) take a very simple form
\begin{eqnarray}
\rho_{ee}&=&\frac{{N}^{2}+N(N+1)\,\gamma_{12}^{2}} 
{1+4\,N(N+1)(1+\gamma_{12}^{2})}\ ,\nonumber \\
\rho_{ss}&=&\frac{N(N+1)(1-\gamma_{12})^{2}} 
{1+4\,N(N+1)(1+\gamma_{12}^{2})}\ ,\nonumber \\
\rho_{aa}&=&\frac{N(N+1)(1+\gamma_{12})^{2}} 
{1+4\,N(N+1)(1+\gamma_{12}^{2})}\ ,\nonumber \\
\rho_{u}&=&\frac{2\,\sqrt{N(N+1)}\,(2N+1)\,\gamma_{12}} 
{1+4\,N(N+1)(1+\gamma_{12}^{2})} .\label{t221}
\end{eqnarray}
The solutions~(\ref{t220}) and~(\ref{t221}) will be used for
calculation of the degree of entanglement present in the system.
For further reference it is important to note that the sum of the
populations $\rho_{ss}+\rho_{aa}$ tends to $0.5$ as the quantity
$N(N+1)(1+\gamma_{12}^{2})$ becomes much greater than one, which means
that for large values of the mean number of photons $N$ one half of
the population goes eventually to the states $|s\rangle$ and
$|a\rangle$.   

\subsection{Nonidentical atoms}
The population distribution is quite different when the atoms are 
nonidentical with $\Delta=(\omega_{2}-\omega_1)/2 \neq 0$. As before 
for the identical atoms, we use the master 
equation (\ref{eq1}) and find four coupled differential equations for 
the density matrix elements with time-dependent coefficients 
oscillating at frequencies exp$(\pm i\Delta t)$ and exp$[\pm
2i(\omega_{s}-\omega_{0})t +\phi_{s}]$. If we tune the squeezed
vacuum field to the middle of the frequency difference between the
atomic frequencies, {\em i.e.}, $\omega_{s} =(\omega_{1}+\omega_{2})/2$, the
terms proportional to exp$[\pm 2i(\omega_{s}-\omega_{0})t +\phi_{s}]$
become stationary in time. None of the other time dependent components
is resonant with the frequency of the squeezed vacuum field.
Consequently, for $\Delta \gg \Gamma$, the
time-dependent components oscillate rapidly in time
and average to zero over long times. Therefore, we can make a
secular approximation in which we ignore the rapidly oscillating
terms and obtain the following equations of motion
\begin{eqnarray}
\dot{\rho}_{ee} &=&
-2\Gamma\left({N}+1\right)\rho_{ee}
+{N}\Gamma \left(\rho_{ss}
+\rho_{aa}\right) +\Gamma_{12}|{M}|\rho_{u}\,  ,\nonumber \\
\dot{\rho}_{ss} &=&
\Gamma \left[{N} -\left(3{N}+1\right)\rho_{ss} -{N}\rho_{aa}
+\rho_{ee}\right] -\Gamma_{12} |{M}|\rho_{u}\, ,\nonumber \\
\dot{\rho}_{aa} &=& \Gamma \left[{N}
-\left(3{N}+1\right)\rho_{aa} -{N}\rho_{ss}
+\rho_{ee}\right] -\Gamma_{12} |{M}|\rho_{u}\, ,\nonumber \\
\dot{\rho}_{u} &=& 2\Gamma_{12}|{M}|-\left(2{N}+1\right)\Gamma \rho_{u} 
-4\Gamma_{12}|{M}| \left(\rho_{ss}+\rho_{aa}\right)\, .\label{eq10}
\end{eqnarray}

The steady-state solutions of equations (\ref{eq10}) are
\begin{eqnarray}
\rho_{ee}
&=&\frac{1}{4}\left\{\frac{\left(2{N}-1\right)}{2{N}+1}+
\frac{1}{\left[\left(2{N}+1\right)^{2}
-4|{M}|^{2}\,\gamma_{12}^{2}\right] }\right\}
\ ,\nonumber \\
\rho_{ss} &=& \rho_{aa} =\frac{1}{4}\left\{ 1-\frac{1}{\left[
\left(2{N}+1\right)^{2}-4|{M}|^{2}\,\gamma_{12}^{2}\right] }\right\}
\ ,\nonumber \\
\rho_{u} &=&
\frac{2|M|\,\gamma_{12}}{\left(2{N}+1\right)
\left[\left(2{N}+1\right)^{2}
-4|{M}|^{2}\,\gamma_{12}^{2}\right]}
\ .\label{t227}
\end{eqnarray}
Equations (\ref{t227}) are quite different from equations~(\ref{t220})
and show that in the case of non-identical atoms
the symmetric and antisymmetric states are equally
populated. This fact will be crucial in the entanglement creation in 
the system and results from the equal damping rates of the symmetric 
and antisymmetric states, as it is seen from equation (\ref{eq10}).
For maximally squeezed vacuum with $|M|=\sqrt{N(N+1)}$ the
solutions~(\ref{t227}) simplify to
\begin{eqnarray}
\rho_{ee}&=&\frac{1}{4}\left\{\frac{2N-1}{2N+1}+\frac{1}
{1+4N(N+1)(1-\gamma_{12}^{2})}\right\}\, ,\nonumber \\
\rho_{ss} &=&\rho_{aa}=\frac{N(N+1)(1-\gamma_{12}^{2})}
{1+4N(N+1)(1-\gamma_{12}^{2})}\, ,\nonumber \\
\rho_{u} &=&
\frac{2\sqrt{N(N+1)}\,\gamma_{12}}{(2N+1)[1+4N(N+1)(1-\gamma_{12}^{2})]}
\, .\label{t227a}
\end{eqnarray}
From equations~(\ref{t227a}) it is evident that for the Dicke model,
for which $\gamma_{12}=1$, the populations of the symmetric 
and asymmetric states both are zero and $\rho_{u}$ tends to unity for
large $N$.
In real situations the separation of the atoms is
nonzero, and we have always $\gamma_{12}<1$, which means that for
$N(N+1)(1-\gamma_{12}^{2})\gg 1$ the populations $\rho_{ss}$ and
$\rho_{aa}$ both approach value $0.25$, {\em i.e.},
$\rho_{ss}+\rho_{aa}\approx 0.5$. That is, for sufficiently large $N$
one half of the population is transferred to the block spanned by the
symmetric and antisymmetric states, similarly to the identical
atoms. There is, however, one essential difference between the identical
and nonidentical atoms, which is the scale for the saturation -- much
slower for nonidentical atoms. For $\gamma_{12}$ very close to unity, 
very large intensities $N$ are required to have 
$N(N+1)(1-\gamma_{12}^{2})\gg 1$, and consequently 
$\rho_{ss}+\rho_{aa}\approx 0.5$. Thus, for small $N$ and 
$\gamma_{12}\neq 1$, the populations $\rho_{ss}$ and $\rho_{aa}$ are 
very small, $\rho_{ss}=\rho_{aa}\approx 0$.
This fact will have important effect on the entanglement creation in 
the system of nonidentical atoms.

\section{Steady state entanglement}

To assess how much entanglement is stored in a given quantum system it
is essential to have appropriate measures of entanglement. A number of
measures have been proposed, which include entanglement of
formation~{\cite{woo}}, entanglement of distillation~\cite{bbpssw96},
relative entropy of entanglement~\cite{vpjk97} and
negativity~\cite{per,horo,zhsl98,vw02}. For pure states, the Bell
states represent 
maximally entangled states, but for mixed states represented by a
density matrix there are some difficulties with ordering the states
according to various entanglement measures; different entanglement
measures can give different orderings of pairs of mixed states and
there is a problem of the definition of the maximally entangled mixed
state~\cite{ih00,wngkmv03}. 

Here we use the concurrence to describe the amount of entanglement created in a
two-atom system by the interaction with the squeezed vacuum. The concurrence
introduced by Wootters~\cite{woo} is defined as 
\begin{equation}
  \label{eq:concurrence}
  {\cal
  C}=\max\left(0,\sqrt{\lambda_{1}}-\sqrt{\lambda_{2}}-\sqrt{\lambda_{3}}
  -\sqrt{\lambda_{4}}\right)\, ,
\end{equation}
where $\{\lambda_{i}\}$ are the the eigenvalues of the matrix
\begin{equation}
  \label{eq:R}
  R=\rho\tilde{\rho}
\end{equation}
with $\tilde{\rho}$ given by
\begin{equation}
  \label{eq:rhot}
  \tilde{\rho}=\sigma_{x}\otimes\sigma_{x}\,\rho^{*}\,
\sigma_{x}\otimes\sigma_{x}\, , 
\end{equation}
$\sigma_{x}$ is the Pauli matrix, and $\rho$ is the density matrix
representing the quantum state.
The range of concurrence is from 0 to 1. For unentangled atoms ${\cal
C}=0$ whereas ${\cal C}=1$ for the maximally entangled atoms.

In the basis~(\ref{eq:basis}) of the product atomic states the density
matrix for two atoms in the squeezed vacuum has in the steady state
the following block form
\begin{equation}
  \label{eq:rho}
  \rho=\left(
    \begin{array}[h]{cccc}
\rho_{11}&\rho_{12}&0&0\\
\rho_{21}&\rho_{22}&0&0\\
0&0&\rho_{33}&\rho_{34}\\
0&0&\rho_{43}&\rho_{44}
    \end{array}\right)
\end{equation}
with the condition $\Tr{\rho}=1$.
The matrix $\tilde{\rho}$, required for calculation of the concurrence, has
the form
\begin{equation}
\label{eq:rhot1}
\tilde{\rho}=\left(
    \begin{array}[h]{cccc}
\rho_{22}&\rho_{12}&0&0\\
\rho_{21}&\rho_{11}&0&0\\
0&0&\rho_{44}&\rho_{34}\\
0&0&\rho_{43}&\rho_{33}
    \end{array}\right)
\end{equation}
and the square roots of the eigenvalues of the matrix $R$ given
by~(\ref{eq:R}) are the following
\begin{eqnarray}
  \label{eq:roots}
\fl \left\{\sqrt{\lambda_{i}}\right\}=&&\left\{
\sqrt{\rho_{11}\rho_{22}}-|\rho_{12}|,
\sqrt{\rho_{11}\rho_{22}}+|\rho_{12}|,\sqrt{\rho_{33}\rho_{44}}-|\rho_{34}|,
\sqrt{\rho_{33}\rho_{44}}+|\rho_{34}|\,\right\} .
\end{eqnarray}
Depending on the particular values of the matrix elements there are
two possibilities for the largest eigenvalue, either the second term or
the fourth term in~(\ref{eq:roots}). 
The concurrence is thus given by
\begin{equation}
  \label{eq:concurrence1}
  {\cal C}=
\max\left\{0,\,{\cal C}_{1},\,{\cal C}_{2}
\right\},
\end{equation}
with
\begin{eqnarray}
  \label{eq:altconc}
  {\cal C}_{1}=2\,(|\rho_{12}|-\sqrt{\rho_{33}\rho_{44}}\,)\ ,\nonumber\\
{\cal C}_{2}=2\,(|\rho_{34}|-\sqrt{\rho_{11}\rho_{22}}\,)\ ,
\end{eqnarray}
and we have two alternative expressions for the concurrence depending
on which of them is positive.

In terms of the collective atomic states
$|g\rangle,|e\rangle,|s\rangle$ and $|a\rangle$, the expressions for the
concurrence~(\ref{eq:altconc}) take the form
\begin{eqnarray}
  \label{eq:concatom}
  {\cal C}_{1}=2\,|\rho_{ge}|-\sqrt{(\rho_{ss}+\rho_{ss})^2
-(\rho_{sa}+\rho_{as})^2}\,)\ ,\nonumber\\ 
{\cal C}_{2}=|\rho_{ss}-\rho_{aa}+\rho_{sa}-\rho_{as}|
-2\,\sqrt{\rho_{gg}\rho_{ee}}\,)\, .
\end{eqnarray}
Having the concurrence expressed in terms of the density matrix 
elements, we can apply the steady-state solutions~(\ref{t220})
or~(\ref{t227}) and obtain analytical results for the stationary
concurrence. We will discuss the results separately for
identical and nonidentical atoms.

\subsection{Identical atoms}

In case of identical atoms the steady-state solutions~(\ref{t220}) are
valid, and we find 
\begin{eqnarray}
  \label{eq:concident}
  {\cal C}_{1}&=&|\rho_{u}|-(\rho_{ss}+\rho_{aa})\nonumber\\
&=&2\,\frac{(2N+1)|M|\,\gamma_{12}-|M|^{2}\gamma_{12}^{2}
-N(N+1)\left[(2N+1)^{2}-4|M|^2\right]} 
{(2N+1)^{4}-4|M|^2\left[(2N+1)^2-\gamma_{12}^{2}\right]}\, . 
\end{eqnarray}
It turns out that ${\cal C}_{2}$ is always negative, so the only
contribution to the concurrence comes from ${\cal C}_{1}$, and then
${\cal C} =\max(0,{\cal C}_{1})$. 
It is clear from~(\ref{eq:concident}) that the perfect
entanglement in the system, {\em i.e.}, the value of concurrence equal
\begin{figure}[htb]
\centering
\includegraphics[width=12cm]{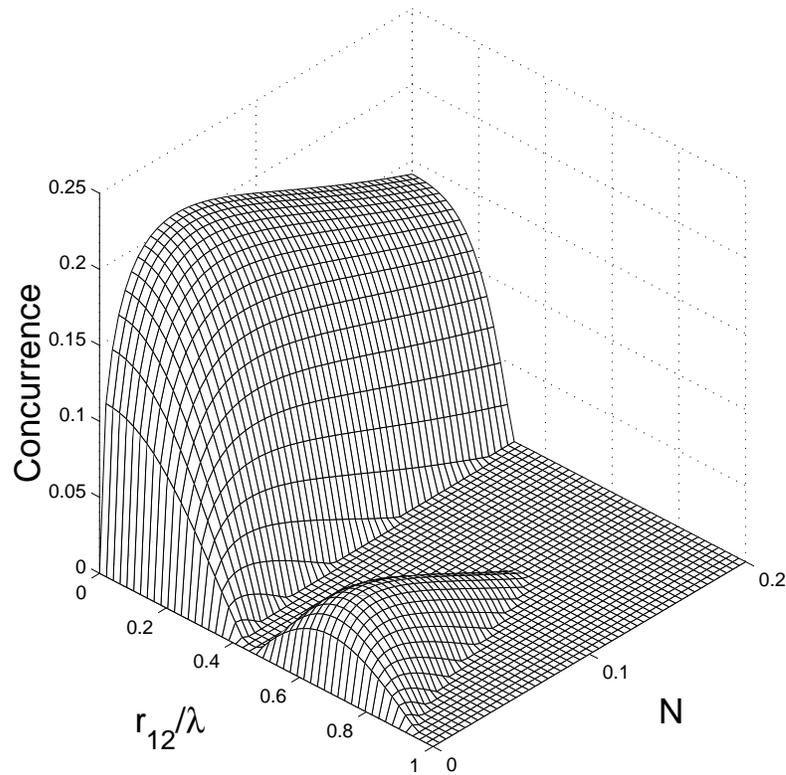}
\caption{Concurrence ${\cal C}$ for two identical atoms as a function of
  the interatomic distance $r_{12}/\lambda$ and the mean number of
  photons $N$ for $|M|=\sqrt{N(N+1)}$}
  \label{fig:1}
\end{figure}
to unity, can be achieved when $|\rho_{u}|=1$ and
$\rho_{ss}+\rho_{aa}=0$. This is generally impossible for identical 
atoms because there is
always some population stored in the states $|s\rangle$ and
$|a\rangle$. However, it is possible to obtain some degree of
entanglement in the system for appropriately chosen values of 
$r_{12}$, $N$ and $|M|$. 
The concurrence, measuring the degree of entanglement, depends on the
interatomic distance through the collective damping parameter
$\gamma_{12}$, and the degree of the two-photon coherences $|M|$. It 
is easily to show from equation (\ref{eq:concident}) that 
there is no entanglement possible for $|M|\le N$, {\em i.e.}, 
for classically correlated fields. For a quantum squeezed field with 
maximum correlations $|M|=\sqrt{N(N+1)}$, the concurrence can be 
written as
\begin{eqnarray}
  \label{eq:concident1}
{\cal C}_{1}&=&2\,\sqrt{N(N+1)}\,\frac{(2N+1)\,\gamma_{12}
-\sqrt{N(N+1)}\,(1+\gamma_{12}^{2})}{1+4\,N(N+1)(1+\gamma_{12}^{2})}\, .
\end{eqnarray}
In this case, ${\cal C}_{1}$ can be positive. To show this, we plot in 
Fig.~\ref{fig:1} the concurrence for two identical atoms in the maximally
squeezed vacuum as a function of the interatomic distance
$r_{12}$ and the mean number of photons $N$. It is evident from
Fig.~\ref{fig:1} that there is a range of values of 
$r_{12}/\lambda$ for which ${\cal C}$ is positive.
The maximum of concurrence is obtained for
$r_{12}\approx 0$ when the atoms are very close to each other. 
The values of concurrence decrease as the
interatomic distance increases and reduces to zero at $r_{12}\approx
\lambda/2$, but we can observe revival of concurrence for longer
interatomic distances, although the next maximum is much weaker.
It is interesting that the
maximum of concurrence appears for not very high values of the mean
number of photons $N<0.1$. It is easy to check,
from~(\ref{eq:concident1}), that for 
$\gamma_{12}=1$ the number of photons
$N_{max}$ for which $C_{1}$ reaches its maximum is equal to
$N_{max}=(\sqrt{(1+\sqrt{2})/2}-1)/2\approx 0.049$. The fact that the
maximum of concurrence appears for moderate values of the mean photon
numbers can be important from the 
experimental point of view as the present sources of squeezed fields
can produce quantum squeezed fields of intensities $N<1$. 

The steady-state entanglement and its presence for only quantum 
squeezed fields of small 
intensities $N$ is associated with nonclassical two-photon
correlations characteristic of the squeezed vacuum field. To show 
this, we introduce a parameter 
\begin{eqnarray}
    TC = \frac{|M|}{N} ,
\end{eqnarray}
which characterises two-photon correlations normalised to the 
intensity of the squeezed field. For classical fields, $|M|\leq N$, 
and then $TC<1$ for all $N$. For a quantum squeezed field with 
$|M|=\sqrt{N(N+1)}$, the parameter becomes
\begin{eqnarray}
    TC = \sqrt{1+\frac{1}{N}} ,
\end{eqnarray}
which is always greater than one. The result is a strong two-photon 
correlation, which is greatest for $N<1$. Thus, the non-classical 
two-photon correlations are significant for $N<1$ and lead to a large 
entanglement in the system.

\subsection{Nonidentical atoms}
In the case of nonidentical atoms with $\Delta\ne 0$, the steady-state
values for the density matrix elements are given in equation~(\ref{t227}).
As above for identical atoms, 
the concurrence~(\ref{eq:concurrence1}) can be 
expressed by the formula~(\ref{eq:altconc}) which, with the
solutions~(\ref{t227}), leads to 
\begin{eqnarray}
  \label{eq:concnonid}
  {\cal C}_{1}=|\rho_{u}|-(\rho_{ss}+\rho_{aa})&=&\frac{2|M||\gamma_{12}|}
{\left(2{N}+1\right)\left[\left(2{N}+1\right)^{2}
-4|{M}|^{2}\,\gamma_{12}^{2}\right]}\nonumber\\
&&-\frac{1}{2}\left\{ 1-\frac{1}{
\left(2{N}+1\right)^{2}-4|M|^{2}\,\gamma_{12}^{2} }\right\}
\end{eqnarray}
Similarly to the case of identical atoms, ${\cal C}_{2}$ is always 
negative. Moreover, ${\cal C}_{1}$ is always negative for $|M|\leq N$
independent of $\gamma_{12}$. Thus, 
entanglement is possible only for quantum squeezed fields which for 
the maximum correlations $|M|=\sqrt{N(N+1)}$ gives
\begin{eqnarray}
  \label{eq:concnonid1}
  {\cal C}_{1}&=&2\,\sqrt{N(N+1)}\,\frac{\gamma_{12}-(2N+1)\sqrt{N(N+1)}\,
(1-\gamma_{12}^{2})}{(2N+1)[1+4N(N+1)(1-\gamma_{12}^{2})]} .
\end{eqnarray}
Equation (\ref{eq:concnonid1}) is significantly different from that 
for identical atoms, equation (\ref{eq:concident1}). For example,
if the atoms are close together, $\gamma_{12}\approx 1$, and then
equation~(\ref{eq:concnonid1}) reduces to
\begin{eqnarray}
  \label{eq:concnonid2}
  {\cal C}_{1}&=&\,\frac{2\,\sqrt{N(N+1)}}{2N+1}\, .
\end{eqnarray}
In this limit the concurrence is always positive, increases with $N$
and approaches unity at a large $N$. This is in contrast to the case of 
identical atoms where values of concurrence are below $0.25$ even for
$\gamma_{12}=1$ and approach zero for large $N$.
This behaviour can be easily explained by the fact
that in the case of nonidentical atoms and $\gamma_{12}=1$ the
population stored in the symmetric and antisymmetric states,
$\rho_{ss}+\rho_{aa}$, is equal to zero. At the same time,  
$\rho_{u}$ tends to unity as $N$ increases giving the maximum 
concurrence ${\cal C}_{1}=1$.

In real situations, we have $\gamma_{12}<1$, and for large $N$ the
concurrence $C_{1}$, given by~(\ref{eq:concnonid1}), goes to zero
similarly to the case of identical 
atoms. The maximum of $C_{1}$ for nonidentical atoms, nonetheless, is
much more pronounced than that for identical atoms.
The real scale of large number of photons is in this case given by
$N(N+1)(1-\gamma_{12}^{2})\gg 1$ rather than by
$N(N+1)(1+\gamma_{12}^{2})\gg 1$ as
it is the case for identical atoms. 

\begin{figure}[htb]
\centering
\includegraphics[width=12cm]{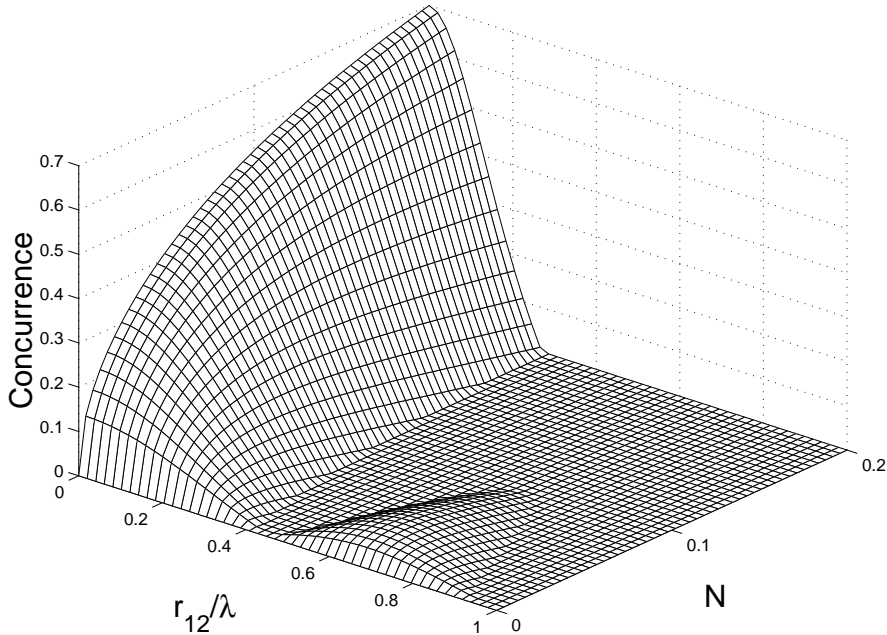}
\caption{Concurrence ${\cal C}$ for two nonidentical atoms as a function of
  the interatomic distance $r_{12}/\lambda$ and the mean number of
  photons $N$ for $|M|=\sqrt{N(N+1)}$}
  \label{fig:2}
\end{figure}

The dependence of the concurrence 
${\cal C}=\max(0,{\cal C}_{1})$ with ${\cal C}_{1}$ given
by formula~(\ref{eq:concnonid1}) on the interatomic distance $r_{12}$
and the mean number of photons of the squeezed field $N$ is shown in
Fig.~\ref{fig:2}.
The dependence on the interatomic distance is similar to that seen for
identical atoms with the revival of concurrence for $r_{12}\approx
3\lambda/4$ and not too large $N$. For $\gamma_{12}\ne 1$ the
dependence on $N$ is also similar to that for identical atoms, except
that the maximum values of concurrence for given $N$ are much higher
than for identical atoms. 
\begin{figure}[htb]
\centering
\includegraphics[width=10cm]{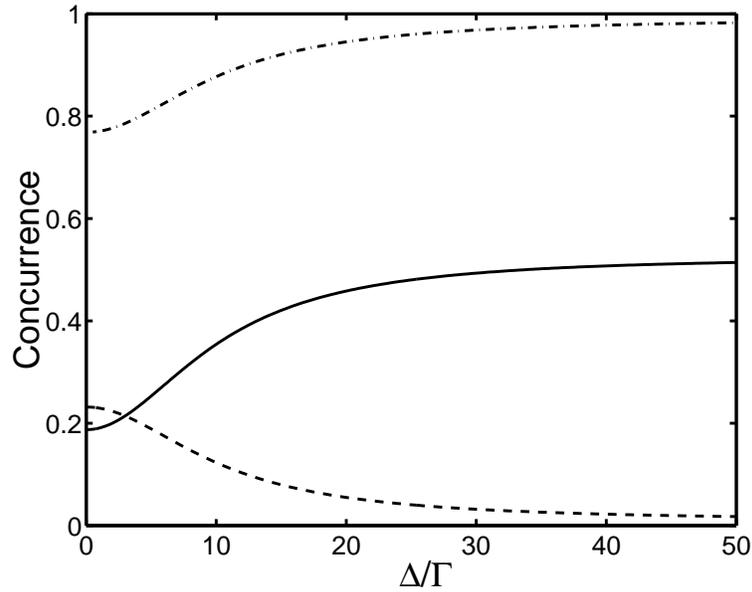}
\caption{Concurrence ${\cal C}$ (solid line), populations
  $\rho_{ss}+\rho_{aa}$ (dashed line), and populations
  $\rho_{gg}+\rho_{ee}$ (dashed-dotted line) for two  atoms as a function of
  $\Delta$ for $r_{12}/\lambda=0.05$ and $N=0.1$ for $|M|=\sqrt{N(N+1)}$}
  \label{fig:3}
\end{figure}
Comparing the solutions for identical and nonidentical atoms indicates
that one reason for higher values of concurrence for nonidentical
as compared to identical atoms interacting with the squeezed vacuum is
the fact that for nonidentical atoms less population remains in the
lower block of the density matrix~(\ref{eq:rho}) represented by states
$|3\rangle$ and  $|4\rangle$ (or $|a\rangle$ and $|s\rangle$) as the
atoms become different, i.e., $\Delta=(\omega_{2}-\omega_{1})/2$
becomes large. To confirm this fact we plot in Fig.~\ref{fig:3} the 
\begin{figure}[htb]
\centering
\includegraphics[width=10cm]{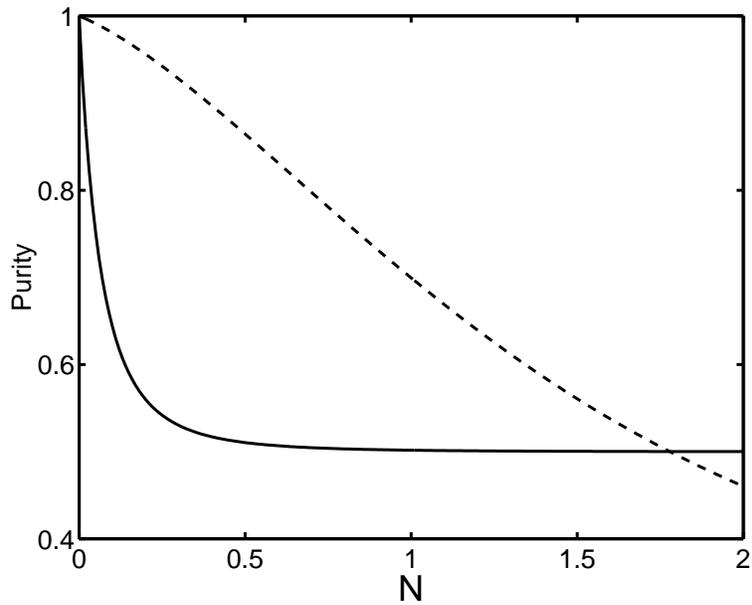}
\caption{Purity $P={\rm Tr}(\rho^{2})$ as a function of the mean
  number of photons $N$ for $r_{12}/\lambda=0.05$ and
  $|M|=\sqrt{N(N+1)}$: identical atoms (solid line), nonidentical
  atoms (dashed line)}
  \label{fig:4}
\end{figure}
concurrence as well as the populations, $\rho_{ss}+\rho_{aa}$ and
$\rho_{gg}+\rho_{ee}$, which are stored in the two blocks of
the density matrix~(\ref{eq:rho}), as a function of $\Delta$. 
Since for identical atoms,
according to~(\ref{t221}), considerable amount of population remains in
the antisymmetric state in contrast to the solutions~(\ref{t227a}) for
nonidentical atoms, it is clear from Fig.~\ref{fig:3} that as the
transition frequencies of the two atoms become more and more different
the population of the antisymmetric state goes down reducing the
total population $\rho_{ss}+\rho_{aa}$ of the lower block and
increasing the total population $\rho_{gg}+\rho_{ee}$ of the upper
block of~(\ref{eq:rho}), which means higher values of concurrence.

Another physical explanation of the origin of the better entanglement 
for non-identical atoms is provided by the observation that the 
stationary state of non-identical atoms, for small $N$ for which
concurrence is maximal, is close to a pure state,  
whilst the stationary state of identical atoms is already far from a pure 
state. This is illustrated in Fig.~\ref{fig:4}, where we plot the purity
measure $P={\rm  Tr}(\rho^{2})=\rho_{gg}^{2}+\rho_{ee}^{2}+\rho_{ss}^{2}
+\rho_{aa}^{2}+|\rho_{u}|^{2}/2$ as    
a function of $N$ for the steady-state  of two identical as well as 
nonidentical atoms.
It is seen that in both cases the purity decreases as the number of
photons increases, but in case of identical atoms the purity goes down
much faster.
\begin{figure}[htb]
\centering
\includegraphics[width=10cm]{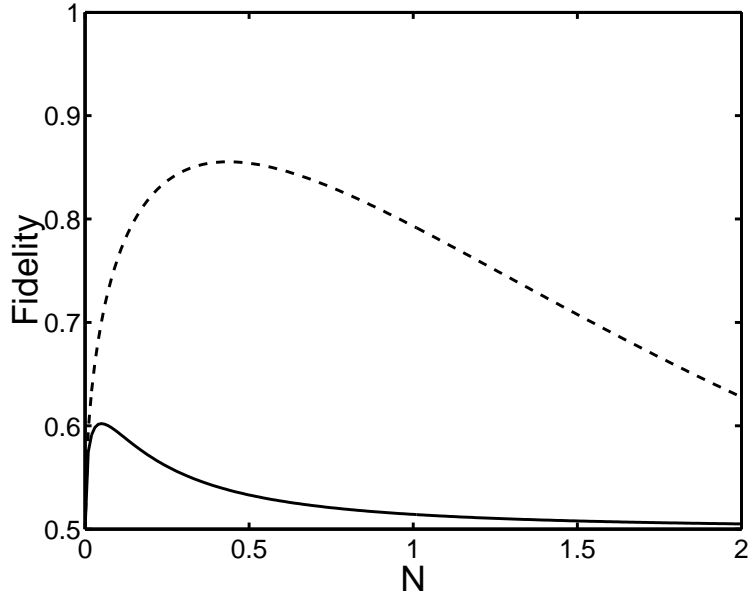}
\caption{Fidelity $F_{+}$ as a function of the mean
  number of photons $N$ for $r_{12}/\lambda=0.05$ and
  $|M|=\sqrt{N(N+1)}$: identical atoms (solid line), nonidentical
  atoms (dashed line)}
  \label{fig:5}
\end{figure}

It should be emphasised, however, that the main source of
entanglement in the system are the nonclassical two-photon correlations
that create two-photon coherences between the states $|g\rangle$ 
and $|e\rangle$. The two-photon coherences are nonzero only when the
squeezing parameter $|M|$ is nonzero. In fact to have entanglement in
the system the squeezed field must represent quantum correlations with
$|M|>N$. There is, moreover, one more necessary condition to have
nonzero $\rho_{u}$, which is a nonzero value of
the collective damping parameter $\gamma_{12}$. The two photon
coherences cause the system to  
decay into entangled states involving the ground state $|g\rangle$ 
and the upper state $|e\rangle$ without any involvement of the 
entangled states $|s\rangle$ and $|a\rangle$. Unfortunately, the
spontaneous emission from the state $|e\rangle$ redistributes some of
the atomic population over the states $|s\rangle$ and $|a\rangle$
limiting in this way the degree of entanglement.

Since the two-photon coherences create superposition (entangled) 
states involving the states $|g\rangle$ and $|e\rangle$, one can ask 
a question: How close is the entangled stationary state of the system 
to one of the maximally entangled Bell states
\begin{eqnarray}
  \label{eq:Bell}
  |\Phi^{+}\rangle&=&\phantom{-}\frac{1}{\sqrt{2}}\left(|g\rangle
    +|e\rangle\right)\, ,\nonumber\\ 
|\Phi^{-}\rangle&=&-\frac{1}{\sqrt{2}}\left(|g\rangle 
-|e\rangle\right)\, .
\end{eqnarray}
To answer this question we calculate the fidelities
\begin{eqnarray}
  \label{eq:fidelity}
  F_{+}&=&\langle\Phi^{+}|\rho|\Phi^{+}\rangle
=\frac{1}{2}\left(\rho_{gg}+\rho_{ee}+\rho_{u}\right)\, ,\nonumber\\
F_{-}&=&\langle\Phi^{-}|\rho|\Phi^{-}\rangle
=\frac{1}{2}\left(\rho_{gg}+\rho_{ee}-\rho_{u}\right)\, .
\end{eqnarray}
The fidelities depend on whether $\rho_{u}$ is positive or negative.
For small interatomic distances $\gamma_{12}$ is positive, and then
the coherence $\rho_{u}$ is positive. Hence, the fidelity $F_{+}$
becomes large while the fidelity $F_{-}$ is small. Thus, we can 
conclude that the stationary state of the system is close to the 
maximally entangled Bell state
$|\Phi_{+}\rangle$. In Fig.~\ref{fig:5} we plot the fidelity
$F_{+}$ as a function of $N$ for identical as well as nonidentical
atoms. Comparing the dependence on $N$ of the fidelity $F_{+}$ and
concurrence $C$ gives us clear evidence that the entanglement in the
system can be related to the Bell state $|\Phi_{+}\rangle$. The state
of the system is of course mixed, but it is closer to the pure Bell
state $|\Phi_{+}\rangle$ for nonidentical atoms than for identical
atoms. 
The entanglement created by the two-photon correlations present in the
squeezed light is limited by the population stored in the
other states. As the number of photons increases, for $\gamma_{12}<1$,
more and more population goes to the states $|s\rangle$ and
$|a\rangle$, and eventually entanglement disappears for both identical
as well as nonidentical atoms. There are optimal values of the mean
number of photons for which the highest possible stationary
entanglement can be obtained.

\section{Conclusions}

In this paper, we have studied analytically the entanglement 
creation in a system of two atoms interacting with a squeezed vacuum 
field. We have demonstrated that nonclassical two-photon correlations 
characteristic of the squeezed field can create a large steady-state 
entanglement in the system.

In our approach we have used master equation to describe a system of
two two-level atoms subjected to a squeezed vacuum field. The two
atoms are coupled to each other via the vacuum field which leads to
the collective damping and collective dipole-dipole type interaction
between the atoms. We have assumed the two atoms to be separated by a
distance $r_{12}$, so the collective parameters depend explicitly on
this distance. Steady-state solutions for the atomic density matrix
have been found for two cases: (i) identical atoms, (ii) nonidentical
atoms. 

We have derived analytical expressions for concurrence which is
used to quantify the amount of entanglement created in the system. 
Our results show that the necessary condition for entanglement are 
nonclassical two-photon correlations of the squeezed field. 
The entanglement also depends on the interatomic separation and the
mean number of photons of the
squeezed vacuum. The necessary condition for entanglement are quantum
correlations of the squeezed field. There is no entanglement for
classically correlated field. We have found that the degree of
entanglement created in the system is a result of competition between
the coherent process of transferring two-photon coherences from the
squeezed vacuum to the atomic system and the incoherent process of
spontaneous emission redistributing atomic population over the states
not involved in the former process.
In particular, we have shown that there is an optimum
value of the mean number of photons for which the concurrence takes
its maximum, and it happens for small number of photons. This is
important from the point of view of practical applications. Moreover,
we have also found that the degree of entanglement obtainable in this
way is much higher when the two atoms are not identical. We have
discussed in detail physical reasons for such behaviour of the
two-atom system.


\end{document}